\documentclass{article}
\usepackage{graphicx}
\usepackage{amsmath}

\begin{document}

\author{I. V. Lerner and I. V. Yurkevich\\
\it School of Physics and Astronomy, University of Birmingham,\\
\it Egbaston, Birmingham B15 2TT, United Kingdom\\\\\small
Proceedings of LXXXI Les Houches School on "Nanoscopic quantum
transport", \\\small Les Houches, France, June 28-July 30, 2004}
\title{ Impurity in the Tomonaga-Luttinger model: a
Functional Integral Approach}

\date{March 15, 2005}
%\address{}

%\frontmatter  \mainmatter
\maketitle
\newcommand\dd{{\operatorname{d}}}
\newcommand\sgn{{\operatorname{sgn}}}
\def\Eq#1{{Eq.~(\ref{#1})}}
\def\Ref#1{(\ref{#1})}
\newcommand\e{{\mathrm e}}
\newcommand\cum[1]{  {\Bigl< \!\! \Bigl< {#1} \Bigr>\!\!\Bigr>}}
\newcommand\vf{v_{_\text{F}}}
\newcommand\pf{p_{_\text{F}}}
\newcommand\ef{{\varepsilon} _{\text{\sc f}}}
\newcommand\zf{z_{_\text{F}}}
\newcommand\zfi[1]{{z_{_\text{F}}}_{#1}}
\newcommand\av[1]{\left<{#1}\right>}

\newpage
\section{Introduction}

The Tomonaga-Luttinger  model  \cite{Tom:50,Lutt:63,ML:65} of
one-dimensional (1D) strongly correlated electrons gives a
striking example of non-Fermi-liquid behaviour
\cite{DzLar:73,EfLar:76,Hal:81}. Since the seminal paper by
Haldane \cite{Hal:81}, where the notion of the Luttinger liquid
(LL) has been coined and fundamentals of a modern bosonization
technique have been formulated, this model and its various
modifications remain at the focus of interest in research in
strongly correlated systems. Without diminishing the role of
standard theoretical methods, such as diagrammatic techniques or
renormalization group approach, one can say that  bosonization
methods make the most powerful theoretical tool for strongly
correlated 1D systems. All these methods, and in particularly the
`canonical' operator bosonization, where the   creation and
annihilation operators of electrons are explicitly represented in
terms of Bose operators and a 4-fermionic Hamiltonian is
eventually diagonalized in the bosonic representation, are
described by Dmitrii Maslov in lecture notes published in this
volume.

The standard operator bosonization is one of the most elegant
methods developed in theoretical physics. However, by its very
formulation it seems both limited to and specific for
one-dimensional physics. A subject of this seminar is to
demonstrate the  existence and usefulness of an alternative way to
bosonize 1D interacting electrons, called the 'functional
bosonization'. It is based on the Hubbard-Stratonovich decoupling
of the four-fermion interaction -- a typical way to ``bosonize" a
fermionic system in higher-dimensional problems. This method
 was elaborated in different ways in a set of papers
 \cite{Fogedby:76,LeeChen:88,fernandez01-1,Yur,GYL}. In this
seminar, we will describe such a functional method in the form
similar to that developed earlier   \cite{Yur,GYL} for the
treatment of the pure LL as well as a single-impurity problem in
the Luttinger model. However, we will employ here the Keldysh
technique (see for reviews  \cite{rammer}) rather than the
Matsubara one used in  \cite{Yur,GYL}.

The essence of the method is to eliminate a mixed fermion-boson
term in the action (resulted from the  Hubbard-Stratonovich
decoupling) by a gauge transformation. Such a procedure  is exact
for the pure 1D Luttinger model and gives a convenient starting
point for including a single backscattering impurity.

The problem of a single impurity in the LL has been actively
investigated by many authors
 \cite{kane,ygm,egger95-1,leclair96-1,oreg,furusaki,fabrizio,delft}.
One of the main results of these considerations
 \cite{kane,ygm,furusaki,fabrizio,delft} was the suppression at low
temperatures of the local density of states (TDoS)  \textit{at the
impurity site} and at a distance from the impurity \cite{GYL} and
the related suppression of the conductance \cite{kane,ygm} and the
X ray edge singularity \cite{furusaki,fabrizio,delft}. Another
prominent result was the dependence of the Friedel oscillations
 \cite{egger95-1,leclair96-1,fernandez01-1} on the distance from
the impurity.

In this seminar, we will  briefly outline only the most important
results for the one-impurity problem in the Tomonaga-Luttinger
model, while giving a slightly more detailed description of the
functional bosonization in the Keldysh technique that was not
previously published. For simplicity, we will only address a
single-mode Tomonaga-Luttinger model, with one species of right-
and left-moving electrons, thus omitting spin indices and
considering eventually the simplest linearized model of a
single-valley parabolic electron band. We will also skip over a
physical introduction, referring the reader to lecture notes by
Dmitrii Maslov published in this volume.

%%%%%%%%%%%%%%%%%%%%%%%%%%%%%%%%%%%%%%%%%%%%%%%%%%%%%%%%%%%%%%%%%%%%%%%%%%%%%%%%%%%%%%%%%%%
\section{Functional integral representation}
%%%%%%%%%%%%%%%%%%%%%%%%%%%%%%%%%%%%%%%%%%%%%%%%%%%%%%%%%%%%%%%%%%%%%%%%%%%%%%%%%%%%%%%%%%%
Within the limitations outlined above, the most generic
Hamiltonian of interacting 1D electrons in the presence of an
external scattering potential $v(x)$ can be written as
\begin{align}
\notag{\hat H}&=\int{\rm
d}x{\hat\psi}^{\dagger}\left[-\frac{\partial_x^2}{2m}
-{\varepsilon} _F + v(x)\right]{\hat
\psi}(x)\\&+\frac{1}{2}\int{\rm d}x{\rm
d}x'{\hat\psi}^{\dagger}(x){\hat\psi}^{\dagger}(x')V_0(x-x'){\hat
\psi}(x'){\hat \psi}(x)\,,
\end{align}
where $V_0(x-x')$ is a bare electron-electron interaction. The
observable quantities to be calculated with this Hamiltonian are
the tunneling density of states (TDoS) and the current as a linear
response to an applied field characterised by the vector potential
$A(x,t)$. We will give the results for the TDoS at the end of this
presentation, but will mainly described techniques for calculating
the current $j(x,t)$. It can be written with the help of the Kubo
formula in terms of the current-current correlation function
thermally averaged ($\left<\ldots\right>_0$) in the equilibrium
Gibbs ensemble:
\begin{align}
j(x,t)=i\int\limits_{-\infty}^{t}{\rm d}t'\int{\rm
d}x'\left<\left[{\hat j}(x,t),{\hat j}(x',t')\right]\right>_0
A(x',t')-\frac{ne^2}{m}A(x,t)\,,
\end{align}
where the current operator is defined in the standard way (in the
units $\hbar=1$ here and elsewhere):
\begin{align}\label{current}
{\hat j}(x,t)=-\frac{ie}{2m}\,{\hat
\psi}^{\dagger}(x,t)\partial_x{\hat \psi}(x,t)+{\rm h.c.}
\end{align}
The   Kubo formula can be rewritten in Keldysh techniques by
defining the contour $C_K$ which runs from $-\infty$ to the
observation time $t$  along the upper bank of the cut  along the
real time axis in the complex time plane and then returns to
$-\infty$  along the lower cut:
\begin{align}\label{lin resp}
j(x,t)=i\int\limits_{C_K}{\rm d}t'\int{\rm d}x'\left< T_K{\hat
j}(x,t){\hat j}(x',t')\right>_0 A(x',t')-\frac{ne^2}{m}A(x,t).
\end{align}
Here we have introduced the chronological operator $T_K$
time-ordering the operators in a descending order along the
contour $C_K$ (with all the times on the lower cut considered as
later times as compared to those on the upper cut).    Although
the  current operators are bilinear in Fermi-operators, for the
future usage we assume in the  standard way that the
Fermi-operators anticommute under $T_K$, assuming additionally
that at equal times ${\hat \psi}^{\dagger}(t)$ is taken at an
infinitesimally later moment then ${\hat \psi}(t)$. Such an
agreement is consistent with the definition of the current
(Eq.\ref{current}). In what follows,  we will choose a more
general contour (the Keldysh contour),   running from minus to
plus infinity above the cut and returning below the cut. In the
operator language, such an extension of the contour corresponds to
the insertion of extra evolution operators: having been not
coupled to the observables (the current operator  in our case),
such evolution operators above and below the cut simply cancel
each other.

Now, any  expression written as the chronological operator average
can be straightforwardly represented as a functional integral over
the fields defined on the contour of the time ordering. We
introduce the action,
\begin{align}
S[{\bar\psi},\psi]=\int\limits_{C_K}{\rm
d}t\left\{i{\bar\psi}\partial_t\psi-H[{\bar\psi},\psi]\right\}\,,
\end{align}
with the last term being the the normal ordered Hamiltonian with
${\hat\psi}^{\dagger}\to{\bar\psi}$ and ${\hat\psi}\to{\psi}$,
where ${\bar\psi}$ and $\psi$ are the anticommuting Fermi fields.
Then, the   the linear response current of Eq.(\ref{lin resp})
takes the form
\begin{align}\label{lin res}
j(x,t)=i\int\limits_{C_K}{\rm d}t'\int{\rm d}x'\left< \widetilde
j(x,t)\widetilde j(x',t')\right> A(x',t')-\frac{ne^2}{m}A(x,t),
\end{align}
where the brackets stand for the functional integral
\begin{align}\label{FI}
\left<\ldots\right>=\int{\cal D}\bar\psi\,{\cal
D}\psi\,(\ldots)\e^{iS[{\bar\psi},\psi]} ,
\end{align}
and the current field is defined by
\begin{align*}
     {\widetilde
j}(x,t)=\frac{e}{2mi}\,{\bar \psi}(x,t)\partial_x \psi(x,t)+{\rm
c.c.}
\end{align*}

\section{Effective action for the Tomonaga - Luttinger Model}
For free electrons, the plane waves basis is natural. Assuming
that the scattering by impurities and electron-electron
interaction involve the energy scales much smaller than the Fermi
energy,  the plane wave basis presents a natural starting point.
Therefore, we may assume that the main contribution to the
functional integral above comes from the fields representing
separately right-moving and left-moving electrons

\begin{align}\label{LR}
    \psi(x,t)\approx\psi_R(x,t)\,e^{ip_Fx}+\psi_L(x,t)\, e^{-ip_Fx}
\end{align}
where $\psi_{R,L}$ are smooth  on the $p_F^{-1}$ scale. Such a
separation is the essence of the Tomonaga-Luttinger model, and
corresponds to the linearization of the initially parabolic
electron band. The right- and left-moving electrons can be
transformed one to another due to backscattering processes. We
shall neglect such processes due to the electron-electron
interaction, thus keeping only the small momentum transfer part of
interaction, $V(q\ll 2p_F)$. This part of interaction is
non-trivial by itself as it breaks the Fermi Liquid theory in 1D.
On the other hand we will keep only backscattering in the impurity
potential $v(x)$ since small-momentum elastic scattering does not
result in any qualitative change in the Luttinger Liquid behavior.

Now we make the substitution (\ref{LR})  neglecting higher order
derivatives of smooth functions and discarding integrals over fast
oscillating terms. After some straightforward manipulations we
come to the action for the Tomonaga-Luttinger model:
\begin{align}\label{TL}
S_{TL}=S_0+S_{\text{int}}.
\end{align}
The first term describes free electrons  in the presence of the
external scattering potential (which can be also changing in
time):
\begin{align}\label{S0}
S_0=\int{\rm d}x{\rm
d}t\,\Psi^{\dagger}(x,t)\left(\begin{array}{cc}i\partial_R &
v(x,t) \\ {\bar v}(x,t) & i\partial_L
\end{array}\right)\Psi(x,t)\,.
\end{align}
We have assumed (here and below) that all the time integrations
are performed along the Keldysh contour, and introduced the
following notations:
\begin{align}\label{a}
\Psi &= \begin{pmatrix}
  \psi_R\cr\psi_L
\end{pmatrix}, &\Psi^{\dagger}&=\begin{pmatrix}
  \bar \psi^R ,& \bar \psi^L \\
\end{pmatrix}\,,&
\partial_{R/L}&\equiv \partial_t\pm v_F\partial_x\,.
\end{align}
The second term in \Eq{TL} gives the interaction part of the
action
\begin{align}\label{int}
S_{\text{int}}&=-\frac{1}{2}\int{\rm d}x{\rm d}x'{\rm d}t\, n(x,t)V_0(x-x')n(x',t)\,,&\Bigl(n &\equiv
\Psi^{\dagger} \Psi \Bigr)\,.
\end{align}

\section {The bosonized action for free electrons}
Before dealing with interacting electrons, we convert the action
\Ref{S0} for  free electrons   in the presence of the impurity
potential $v$
  into the action in terms of bosonic fields.

  Expanding Eq.\ (\ref{S0}) in powers of
  $v$, we obtain the  partition function by integrating
over the fermion fields with the help of the Wick theorem:
\begin{align}\label{Z0}
\!\!\!\!Z_0=\sum_{n=0}^{\infty}\frac{(-1)^n}{(n!)^2}\int\!\!{\rm
d}^n z\,{\rm d}^n z'  \det g_L(z_i,z'_j)\det
g_R(z'_i,z_j)\prod_{k=1}^{n}v(z_k){\bar v}(z'_k) .
\end{align}
We have introduced the notation $z=(x,t)$ and defined the Green
functions of left- and right-moving electrons in the standard way
(suppressing indices $L,R$):
\begin{align}\label{Keldysh g}
    g\equiv -\rm i\left<\psi(z)\bar \psi(z')\right>= \begin{pmatrix}
     g^{++} & g^<\\
      g^> & g^{--}
    \end{pmatrix}\,.
\end{align}
The matrix structure is defined by the position of  the time
arguments of $\psi$ and $\bar \psi$: in $g^{++}$ and $g^{--}$ both
of them are, respectively, on the upper or lower branches of the
contour, and in $g^{<}(g^>)$ the first (second) argument is on the
upper branch while the second (first) is on the lower. We should
remind  that the functional average \Ref{FI}, invoked in the
definition \Ref{Keldysh g}, automatically arranges for the time
ordering along the Keldysh contour. The Green function obeys the
equation
\begin{align}
{\rm i}\partial_{\eta}g_{\eta}(z,z')&=\delta(x-x')\delta({
t,t'}), & (\eta&\equiv R,L)\,.
\end{align}
{Here $\delta(t,t')\equiv \delta(t-t')$ when both arguments are on
the same branch of the  contour and $\delta(t,t')=0$ otherwise;
$\partial _\eta$ is defined in Eq.\ (\ref{a}). } Explicitly
\begin{align}\label{g}
g^>_ {R/L}(z,z')=-\frac{T}{2v_F} \frac{1}{\sinh\pi T\left[(t-t'{
-\rm i0})\mp\frac{x-x'}{v_F}\right]}
\end{align}
{while $g^<$ is different by the sign of $\rm i0$}.
 The components of $g$  obey the usual relation $g^{++}+g^{--}=
g^<+g^>$. In what follows, we will perform the standard Keldysh
rotation, reducing all the appropriate matrices to the triangular
form with the Keldysh component ($g^K=g^<{\bf +}g^>$) in the upper
right corner, and the retarded and advanced components
($g^{r}=(g^a)^* =g^{++}-g^<$) on the main diagonal (see, e.g.,
ref.~ \cite{rammer}). We shall assume that the time arguments
belonging to the upper and lower branch of the contour have,
respectively, positive and negative infinitesimal shift into the
complex plane. Note finally that in the equilibrium case presented
here  the (Fourier transform of the) Keldysh component is related
to the (Fourier transforms of the) retarded and advanced ones via
the Fermi distribution function $f_{\varepsilon} (T) $ as follows:
\begin{align}\label{Kra}
    g^K({{\varepsilon} } )= \bigl(1-2f_{{{\varepsilon} } } (T)\bigr)
    \bigl(g^r({{\varepsilon} } )
    - g^a({{\varepsilon} } )\bigr)\,.
\end{align}
The same relations are valid for other Green's functions to be
considered so that, wherever this does not involve an ambiguity,
we will give only the retarded/advanced components in the explicit
form implying that the expression like \Eq{Kra} is valid for the
Keldysh component.

 To calculate the partition function  \Ref{Z0} we use
the Cauchy identity  \cite{zinnjustin}:
\begin{align}
\det\frac{1}{\sinh (z_i-z_j')}= (-1)^{
n+1}\,\,\frac{\displaystyle\prod_{i<j}\sinh (z_i-z_j)\sinh
(z_i'-z_j')} {\displaystyle\prod_{i,j}\sinh (z_i-z_j')}\,.
\end{align}
 It reduces \Eq{Z0} for the partition
function to the following one:
\begin{align}
Z_0=\sum_{n=0}^{\infty}
\frac{(-1)^n}{(n!)^2}\!\left(\frac{T}{2v_F}\right)^{\!\!2n}\!\!
\!\int\!\!{\rm d}^n\! z\,{\rm d}^n\!
z'\left[\prod_{k=1}^{n}v(z_k){\bar
v}(z'_k)\right]\frac{\prod_{i<j}s(z_i\!-\!z_j)
s(z'_i\!-\!z'_j)}{\prod_{i,j}s(z_i\!-\!z'_j)}
\end{align}
with
\begin{align}
s(z-z')\equiv \sinh\pi
T\!\left(t-t'-\frac{x-x'}{v_F}\right)\,\sinh\pi
T\!\left(t-t'+\frac{x-x'}{v_F}\right)
\end{align}
Introducing the bosonic Green function
\begin{align}\label{GF0}
iG_0(z-z')=-\ln s(z-z')
\end{align}
whose retarded and advanced components are Fourier transform of
\begin{align}\label{G}
G_0^{r/a}(q,{\omega} )=\frac{4\pi v_F}{\omega^2_\pm-v_F^2q^2 }
\,,\qquad {\omega} _\pm\equiv {\omega} \pm i0\,,
\end{align}
one can write
\begin{align}\label{E}
Z_0&=\sum_{n=0}^{\infty}\frac{
(-1)^n}{(n!)^2}\!\left(\frac{T}{2v_F}\right)^{\!\!2n}\!\!
\!\int\!\!{\,\prod_{k=1}^{n}\rm d} z_k\,{\rm d}
z'_k\,v(z_k){\bar v}(z'_k) \notag\\
&\times\exp\Bigl[{-i\sum\limits_{i<j}\left[G_0(z_i-z_j)+
G_0(z'_i-z'_j)\right]+i\sum\limits_{i,j}G_0(z_i-z'_j)}\Bigr]\,.
\end{align}
This is a partition function of the Coulomb gas with the
logarithmic interaction and it can be represented as the
functional integral over the bosonic field $\varphi(x,t)$:
\begin{align}\label{Z}
Z_0=\int\!{\cal D}\!\varphi\, \exp\left\{iS_0[\varphi]+{\alpha}
\!\!\int\!{\rm d}z\!\left[ve^{-i\varphi}+{\rm
c.c.}\right]\right\}\,.
\end{align}
Here the free bosonic action $S_0[\varphi]$ is defined in terms of
the Green's function $G_0$ of \Eq{GF0} as follows
\begin{align}\label{S00}
   S_0[\varphi]
   = \frac{1}{2}\int\!\!\dd z\dd z'\varphi(z)
   G_0^{-1}(z-z')\varphi(z')\,,
\end{align}
The constant ${\alpha} $ in \Eq Z absorbs an ill-defined value of
$G_0(x,t)$ at $x\!=\!0, t\!=\!0$. We use the ultra-violet cutoff
which corresponds to the  scale of order  $\varepsilon_F$:
$$
 {\alpha} =\frac{T}{2v_F}\,
 \e^{\frac{i}{2}G_0(0)}\simeq\frac{\varepsilon_F}{2\pi
 v_F}\,.
$$
Thus we have cast the original free fermion problem into that of
interacting bosons, represented by the partition function \Ref Z.
The interaction between bosons, i.e.\ the second term in the
exponent in \Eq Z, comes from the back\-scattering impurity term
in the original fermionic problem. The   Gaussian action for
noninteracting bosons, \Eq{S00}, can be explicitly written via $x$
and $t$  in the standard form
\begin{align}\label{free bosons}
S_0\left[\varphi\right]=\frac{1}{8\pi v_F}\int\!{\rm d}t\dd x
\left[\left(\partial_t\varphi\right)^2
-v_F^2\left(\partial_x\varphi\right)^2\right]\,,
\end{align}
although \Eq{S00} is no less convenient, especially for the
generalization for the interaction. The main advantage of the
bosonization, either in the standard or functional form, is that
including the quadric electron-electron interaction does not
substantially change the free  action.

\section{Gauging out the interaction}

 The first step in including  the interaction term \Ref{int} into the
bosonization scheme is to perform the Hubbard-Stratonovich
transformation which can be symbolically written as
\begin{align}\label{HS}
\exp\left\{iS_{\text{int}}[\bar\psi,\psi]\right\} =\int{\cal
D}\phi\exp\left\{-\frac{i}{2}\phi V_0^{-1}\phi +\phi
\Psi^{\dagger}\Psi\right\}.
\end{align}
Note that the auxiliary `Hubbard-Stratonovich' (HS)
 bosonic field $\phi$ here is different from the field
$\varphi$ in Eqs.\ \Ref Z and \Ref {free bosons}. Substituting
this representation into the full action \Ref{TL}, we bring the
partition function to the following form:
\begin{align}\label{Z1}
Z=\int{\cal D}\phi\,{\cal D}\Psi^\dagger {\cal D}\Psi\,\e^{i
S_0[\phi]+iS[\Psi,\phi]}\,.
\end{align}
The action $S[\Psi,\phi]$ for fermions interacting with  the HS
field is given by
\begin{align}\label{mixed}
S[\Psi,\phi]=\int{\rm d}z\,\Psi^{\dagger}(z)
\left(\begin{array}{cc}i\partial_R{ -i\phi} & v(z) \\ {\bar v}(z)
& i\partial_L{-i\phi}
\end{array}\right)\Psi(x,t).
\end{align}
To cast  this integral into the form identical to that of the
previous section we apply the local gauge transformation,
\begin{align}\label{gauge}
\psi_{\eta}(z)\equiv \psi_{\eta}(x,t)\to\psi_{\eta}(x,t)\,
\e^{i\theta_{\eta}(x,t)} \quad {\rm with} \quad {
i}\partial_{\eta}\theta_{\eta}(x,t)=\phi(x,t)\,,
\end{align}
which removes the bosonic field $\phi$ from the diagonal part of
the action \Ref{mixed} but at a cost: the off-diagonal terms are
rotated with the factors $\e^{\pm i{\theta} }$, and the Jacobian
of the transformation $J$ changes the quadratic in $\phi$ part of
the action.

It is shown in  Appendix  A that the Jacobian $J$ of the gauge
transformation \Ref{gauge} can be represented as
\begin{align}\label{j}
\ln J[\phi]={\bf -}\frac{i}{2} \int \dd z\dd z'\phi(z)\,\Pi(z,z')\,\phi(z')\,.
\end{align}
The polarization operator $\Pi$ is given in the random phase
approximation (RPA) by
\begin{align}\label{P}
\Pi&=\sum_{\eta=R,L}\Pi_{\eta}, &
\Pi_{\eta}(z-z')&=ig_{\eta}(z-z')g_{\eta}(z'-z)\,,
\end{align}
where $g_{\eta}(x-x',t-t')$ is the free electron Green
 function given by \Eq g.
It is well known that the RPA is exact for the Luttinger Liquid
 \cite{DzLar:73}. Note that we give in  Appendix A a very simple
and straightforward proof of this.

 The
quadratic in $\phi$ contribution of the Jacobian, \Eq j, should be
added to the quadratic term in \Eq{HS}. This results in the free
bosonic action with the kernel corresponding to the screened
interaction:
\begin{align}\label{B}
S[\phi]&={\bf -}\frac{1}{2}\int\!\dd z\dd z'\phi(z)
V^{-1}(z-z')\phi(z'),& V^{-1}&=V_0^{-1}+\Pi\,.
\end{align}
The above expression should be understood in the operator sense:
$V$ and $\Pi$ are the operators whose kernels are defined with the
appropriate time and spatial dependence on the Keldysh contour.

Using the expressions for the retarded and advanced components of
$g_\eta$ (with $\varepsilon_{\pm}=\varepsilon\pm i0$),
\begin{align}\label{gf}
g_R^{r/a}(q,\varepsilon)&=\left(\varepsilon_{\pm}-v_Fq\right)^{-1}\,,
& g_L^{r/a}(q,\varepsilon)&=
\left(\varepsilon_{\pm}+v_Fq\right)^{-1}\,,
\end{align}
one  finds the appropriate components of the polarisation
operator,
\begin{align}\notag
\Pi_R^{r/a}(q,\omega)&=-\frac{1}{2\pi}\frac{v_Fq}{\omega_{\pm}-v_Fq},
&
\Pi_L^{r/a}(q,\omega)&=\frac{1}{2\pi}\frac{v_Fq}{\omega_{\pm}+v_Fq},
 \end{align}
and thus the total  polarisation operator as
\begin{align}\label{RPA}
\Pi^{r/a}(q,\omega)=
-\frac{1}{\pi}\frac{v_Fq^2}{\left(\omega_{\pm}\right)^2-v^2_Fq^2},
\end{align}
Assuming that the Fourier transform of the forward-scattering pair
interaction  only weakly depends on momentum, i.e.\ ${ V}_0(q\!\ll
\!2\pf)\approx {\rm const}\equiv  \bar V$, and substituting
\Eq{RPA} into the free bosonic action \Ref B, one finds the
components of the free HS bosonic propagator as follows:
\begin{align}
V^{r/a}(q,\omega)&=
\frac{\omega_{\pm}^2-v^2_Fq^2}{\omega_{\pm}^2-v^2q^2}\,\bar V \,,
\notag\\[-6pt]\label{PP}\\[-6pt]
V^K(q,\omega)&=\tanh\left(\frac{\omega}{2T}\right)\left[V^R(q,\omega)-
V^A(q,\omega)\right].\notag
\end{align}
Here we introduced the renormalized velocity $v$ which defines the
effective coupling constant g:
\begin{align}\label{coupling}
v^2&\equiv v_F^2+\frac{v_F\bar V }{\pi}\,,& g&\equiv \frac{v_F}{v}
\,.
\end{align}

Therefore, the gauge transformation \Ref{gauge} reduces the action
in \Ref{Z1} to
\begin{align}
    \label{s}
    S=S[{\theta} ]+S[\Psi^\dagger,\Psi; {\theta} ]\,.
\end{align}
Its fermionic part is given by
\begin{align}\label{f}
S[\Psi^\dagger,\Psi; {\theta} ]=\int \!\dd z
\Psi^{\dagger}(z)\left(\begin{array}{cc}i\partial_R & v\,
\e^{-i\theta} \\ {\bar v}\,\e^{i\theta} & i\partial_L
\end{array}\right)\Psi(z)
\end{align}
with $\theta=\theta_R-\theta_L$, while its bosonic part $S[{\theta} ]$
  is defined via the field $\phi$ by Eqs.\ \Ref B, \Ref {PP} and \Ref{coupling}.
It is convenient to write it explicitly as an integral over the
field $\theta$, which is straightforward since ${\theta} $ is
linearly related to the field $\phi$ as in \Eq{gauge}. Thus we
arrive at the following explicit expression for  $ S[{\theta} ]$
in \Eq s:
\begin{align}\label{b}
S[{\theta} ]=\frac{1}{2}\int\dd z\,\dd z'\theta(z)
G_B^{-1}(z-z')\theta(z')\,.
\end{align}
The Gaussian kernel of this interaction, $G_B$, can be represented
as
\begin{align}
    G_B=G-G_0\,,\label{GGG}
\end{align}
where $G_0$ is defined by Eqs.\ \Ref{GF0} and \Ref G, while  $G$
has the standard triangular matrix structure in the Keldysh space,
with the Fourier transform of its retarded/advanced component
given by
 \begin{align}
    \label{FT}
    G ^{r/a}(q,{\omega} )=\frac{4\pi v_F }{\omega_\pm^2-v ^2q^2
    }\,,
\end{align}
i.e.\ it differs from $G_0$ only by substituting $v$ for $v_F$ in
the denominator. Finally, the expressions for fermion density and
current in new variables become
\begin{equation}\label{jn}
     n=
{\bar\psi}_R\psi_R+{\bar\psi}_L\psi_L
+\frac{1}{2\pi}\partial_x\theta, \,\,\,
j=v_F\left[{\bar\psi}_R\psi_R-{\bar\psi}_L\psi_L\right]
-\frac{1}{2\pi}\partial_t\theta
\end{equation}
 which can be seen by keeping
source terms coupled to original fermionic fields when calculating
the Jacobian.

The effective action \Ref s is quadratic in fermionic fields which
can now be easily integrated out. Before doing so, let us stress
that the representation of Eqs. \Ref s--\Ref b seems to be more
convenient for some problems  than the fully bosonized action.
 To perform the fermionic integration, we note that the fermionic
 part of the action, \Eq f, differs from that for the free
 electrons, \Eq {S0}, only by the substitution $v\to v\,\e^{-i\theta}
 $. Therefore, repeating the same procedure as in the previous
 section, we represent this part of the action with the help of
 the bosonic field $\varphi$ so that
 the full action in \Eq s goes (in symbolical notations) to
\begin{align}\notag
S[\varphi, {\theta} ]=\frac{1}{2}\theta
G_B^{-1}\theta+\frac{1}{2}\varphi G_0^{-1}\varphi + {\alpha}
\left[ve^{-i\left(\varphi+\theta\right)}+{\rm c.c.}\right].
\end{align}
Introducing the new bosonic field, $\Theta\equiv \theta+\varphi$,
and noting again the relation \Ref {GGG} we arrive at the standard
fully bosonized action,
\begin{align}\label{A}
S[{\Theta} ]=\frac{1}{2}\int\!\!\dd z\,\dd z'\Theta(z)
G^{-1}(z-z')\Theta(z') +{\alpha}\!\!\int \!\!\dd z
\left[v(z)\,\e^{-i\Theta(z)}+{\rm c.c.}\right]\,,
\end{align}
with \Eq{jn} transforming to
\begin{equation}\label{Sts}
    n=n_0+\frac{1}{2\pi}\partial_x\Theta,\quad
j=-\frac{1}{2\pi}\partial_t\Theta\,.
\end{equation}

\section{Tunnelling density of states near a single impurity}

As an application of the formalism developed above we will
calculate the density of states in the vicinity of a single
impurity in the Luttinger Liquid characterised by the following
local time-independent potential:
\begin{align}\label{v}
v(x)=\lambda v_F\delta(x)
\end{align}
where $\lambda$ is the dimensionless impurity strength. The
potential \Ref v should be substituted into the action \Ref A.
Then one integrates out the fields with $x\neq 0$ which results in
the local action in terms of $\Theta(t)\equiv \Theta(x\!=\!0,t)$:
\begin{align}
S_{\text{imp}}=\frac{1}{2}\int\limits_{C_K}\!{\rm d}t{\rm d}t'
 \Theta(t) G_{\text{imp}}^{-1}(t-t')\Theta(t')
+2{ i}{\alpha}{\lambda}v_F \int\limits_{C_K}\!{\rm d}t\cos {
\Theta(t)}
\end{align}
where the Fourier transform of the retarded/advanced components of
the inverse Gaussian kernel, $G_{\text{imp}}(t)$, are obtained by
integrating the Green's functions of \Eq{FT} over all momenta:
\begin{align}
G_{\text{imp}}^{r/a}(\omega)=\int\frac{{\rm d}q}{2\pi}\frac{4\pi
v_F}{\omega^2_\pm-v^2q^2}=-ig\frac{2\pi}{\omega \pm i0}\,.
\end{align}

We now employ the self-consistent harmonic approximation (see,
e.g.,  \cite{saito,gnt}), i.e.\ substitute the impurity $\cos
{\Theta} $ term with the quadratic one:
\begin{align}\notag
i2av_F\lambda\int{\rm d}t\,\cos\Theta(t) \to-
\frac{i}{2}\Lambda\int{\rm d}t\,\Theta^2(t)
\end{align}
The coefficient $\Lambda$ is to be found from the condition that
this substitution is optimal,
\begin{align}\label{scha}
\frac{\partial}{\partial\Lambda}\left[2av_F\lambda\left<\cos\Theta\right>
- \frac{1}{2}\Lambda\left<\Theta^2\right>\right]=0\,,
\end{align}
where the averages are taken with the  effective action
symbolically represented as
\begin{align}\label{eff act}
S_{{\text{eff}}}=\frac{1}{2}\Theta
\left(G_{\text{imp}}^{-1}+\Lambda\right)\Theta\,.
\end{align}
Solving self-consistently \Eq{scha} with the action \Ref{eff act}
(which involves preserving the proper analytical structure, with
${\Lambda} $ being of the standard matrix structure in the Keldysh
space) one finds with the logarithmic accuracy:
\begin{align}\label{Lambda}
    \Lambda^{r/a}=\pm\varepsilon_F\left(\frac{av_F}{\varepsilon_F}
\lambda\right)^{\frac{1}{1-g}}\simeq
\pm\varepsilon_F\,\lambda^{\frac{1}{1-g}}\,.
\end{align}

  Including the source terms of \Eq{Sts} does not cause any
principal difficulties but makes the transformations more
cumbersome. Therefore, we do not describe such transformations in
the framework of this seminar presentation. Instead, we simply
present the results for the tunnelling density of states
[following from the long time asymptotics of the full electron
Green function calculated in the SCHA with the bosonized action
\Ref s] as a function of the distance from the impurity. We refer
the reader interested in detail to our previous publication
 \cite{GYL} (albeit for the Matsubara rather than the Keldysh
functional integrals).

\begin{figure}[t]
\includegraphics[width=0.75 \textwidth,angle=-90]{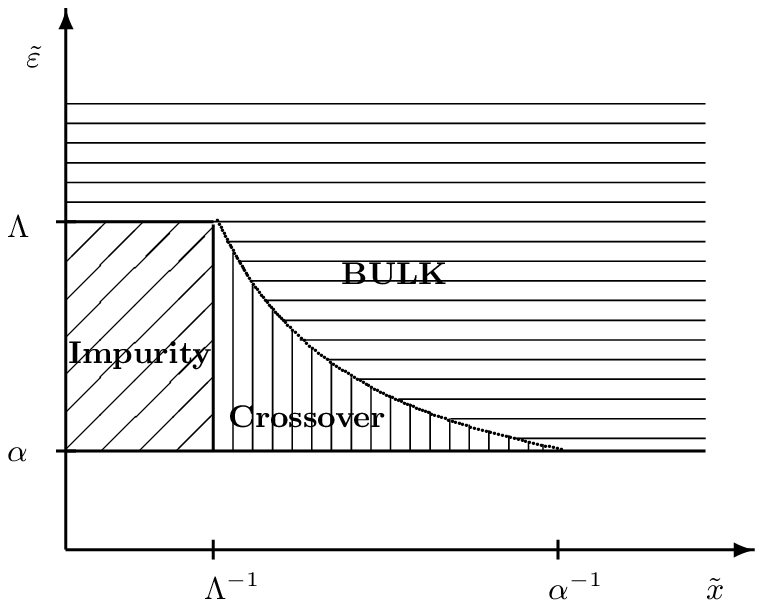}
\end{figure}

We present the explicit expressions for the TDoS smoothed over the
length scale much larger than $\pf^{-1}$ in three different
regions:

\begin{align}
\!\!\!\nu(x,\varepsilon)\sim\left\{\!\!\!\!\!
\begin{array}{cllr}
&\displaystyle \tilde\varepsilon^{\frac{1}{g}-1}
\Lambda^{-\frac{1}{2}\left(\frac{1}{g}-g\right)},
& \tilde{x}\ll\Lambda^{-1}\ll\tilde\varepsilon^{-1}&(a) \\
& \phantom{0} & \phantom{0} \\
& \displaystyle
\tilde\varepsilon^{\frac{1}{g}-1}\tilde{x}^{\frac{1}{2}
\left(\frac{1}{g}-g\right)}, &
\Lambda^{-1}\ll\tilde{x}\ll\tilde\varepsilon^{-1} &(b)\\
& \phantom{0} & \phantom{0} \\
& \displaystyle
\tilde\varepsilon^{\frac{1}{2}\left(\frac{1}{g}+g\right)-1},
 & \!\!\!\!\!\!\!\!\!\min(\tilde{x},\,\Lambda^{-1})\gg\tilde\varepsilon^{-1}&(c)
\end{array}
\right. \label{TDoS}
\end{align}
The regions of different behavior of the tunneling density of
states are sketched in the figure. There and in \Eq{TDoS},
$\tilde{x}\equiv g\pf |x|,\,\, \tilde{{\varepsilon} }\equiv
{\varepsilon} /\ef $,  and the renormalized impurity strength
${\Lambda}\equiv |\Lambda^{r,a}|$ is given by \Eq{Lambda}.
 Equation (\ref{TDoS}a) describes TDoS in the vicinity of
impurity, in full correspondence with the original results of Kane
and Fisher  \cite{kane} obtained for the TDoS at $x=0$, i.e.\
exactly at the impurity. In addition, we have established here the
TDoS dependence   on the impurity strength ${\Lambda}\equiv
\lambda^{\frac{1}{1-g}}$. The region of applicability of
Eq.~(\ref{TDoS}a) corresponds to the diagonally hatched region in
Fig.~1. Equation (\ref{TDoS}c) gives the TDoS at very large
distances from the impurity. As expected, it coincides with a
well-known result for the TDoS in the homogenous Luttinger liquid.
Its region of applicability is horizontally hatched in Fig.~1. In
the intermediate region, vertically hatched in Fig.~1, the TDoS
depends both on the energy and the distance from the impurity.
This analytic dependence given by Eq.~(\ref{TDoS}b) describes  the
crossover from the impurity-induced dip in the TDoS to the bulk
behavior. Finally, the unhatched region for $\tilde {\varepsilon}
<{\alpha} $ corresponds to small energies, ${\varepsilon} <  T$,
where the energy dependence saturates (by ${\varepsilon} \to T$)
in all the three lines of \Eq{TDoS}.

In conclusion, we have demonstrated how to develop the formalism
of bosonization based on the functional integral representation of
 observable quantities within the Keldysh formalism.  We have
 derived in this way the fully bosonized action for the interacting
 electrons in the presence of the scattering potential given by \Eq A,
 and illustrated its usage on the example of the TDoS on a single
 impurity, \Eq {TDoS}. Let us stress
finally that the intermediate representation of Eqs. \Ref s--\Ref
b, which still contains the part quadratic in fermion fields
appears to be more convenient for some problems than the fully
bosonized action.

{\bf Acknowledgements.} We gratefully acknowledge support by the
EPSRC grant  GR/R95432.
\appendix\section{Jacobian of the gauge transformation}

The Jacobian of the gauge transformation (\ref{gauge}) can be
defined  as $J=J_RJ_L$ with
\begin{align}\label{J}
    J_\eta[\phi]=\frac{\int{\cal D}\psi\,\e^{-\int\dd z\,
    \bar{\psi}[\partial_\eta-i\phi]\psi}}
    {\int{\cal D}\psi\,\e^{-\int\dd z\,\bar{\psi}\partial_\eta\psi}}
    =  \e^{{\rm Tr}\ln\left[1-ig_\eta \phi\right]}\,,
\end{align}
where the Green functions of non-interacting right- or left-moving
electrons, obeying $\partial_{\eta}g_{\eta}=\hat I$, are given  by
\Eq{g}. Note that in the matrix components  $g^<(t)$ and $g^>(t)$
the time argument should be understood, respectively, as $t\pm
i0$. The exponent in \Eq J can be represented as infinite series
in the HS-field $\phi$:
\begin{align}
    \ln J_\eta[\phi]=-\sum_{n=1}^{\infty}\frac{i^n}{n!}\operatorname{
    Tr}\left[g_\eta\phi\right]^n
\end{align}
The $n$-th order term in $\phi$ is proportional to the loop
$\Gamma_n^{\eta}$  with $n$ external lines corresponding to
$\phi$'s, each loop being built of the $n$ Green functions
$g_\eta$:
\begin{align}\label{integral}
   \operatorname{
    Tr}\left[g_\eta\phi\right]^n
     \propto\int
     \biggl[\,\prod_{i=1}^n{\rm d}z_i\, \phi(z_i)\biggr]\,
     \Gamma_n\left(z_1;...;z_n\right)\,,
\end{align}
where the $n$-th order vertex  is given by
\begin{align}\notag
\Gamma_n\left(z_1;...;z_n\right)&=\prod_{i=1}^n\,g_\eta\left(z_i;z_{i+1}\right),
&  \bigl( z_{n+1}&= z_1\bigr)\,.
\end{align}
Introducing the new variables $\xi=\e^{2\pi
T\left[\frac{x}{v_F}-t\right]}$, one represents the Green
functions of \Eq g as
\begin{align}\notag
g_R(\xi_1,\xi_2)\propto \frac{\sqrt{\xi_1\xi_2}}{\xi_1-\xi_2}
\end{align}
so that the vertex becomes
\begin{align}
\Gamma_n\left(\xi_1;...;\xi_n\right)
&\propto\gamma_n\left(\xi_1;...;\xi_n\right)\prod_{i=1}^n\xi_i, &
\gamma_n&=\prod_{i=1}^{n} \frac{1}{\xi_{i}-\xi_{i+1}}\,,
\end{align}
with the boundary condition $\xi_{n+1}=\xi_1.$ Since only the
symmetric part of the vertex contributes to the integral
\Ref{integral}, we may symmetrize $\gamma_n$:
\begin{align}
\gamma_n\longmapsto\frac{{\cal
A}_N\left(\xi_1;...;\xi_n\right)}{\prod_{i<j}\left(\xi_i-\xi_j\right)}\,,
\end{align}
where ${\cal A}_N$ is an absolutely antisymmetric polynomial of
the $N$-th order which depends on $n$ variables. A simple power
counting gives $N=n(n-3)/2$ while the possible minimal order of
such a polynomial is $N_{min}=n(n+1)/2$. This is not
self-contradictory for $n=1$ and $n=2$ only. All other terms must
therefore be equal to zero. The term with $n=1$ is cancelled due
to electroneutrality. The only non-vanishing vertex then is the
one with two legs:
\begin{align}
\ln J_\eta\left[\phi\right]=\frac{i}{2}\int\!{\rm d}z{\rm
d}z'\,\phi(z)\Pi_\eta(z;z')\phi(z'),
\end{align}
where
\begin{align}
\Pi_{\eta}(z;z')=ig_{\eta}(z-z')g_{\eta}(z'-z)\,.
\end{align}

%\bibliography{ag01,com,my}\end{document}

{

\end{document}
This analytic dependence, as well as the general representation of
the Green function in the presence of the impurity, are the main
results of the paper.